
\font\mybb=msbm10 at 12pt
\def\bb#1{\hbox{\mybb#1}}
\def\Z {\bb{Z}}
\def\R {\bb{R}}

\def\j { j}
\def\tfrac#1#2{{\textstyle{#1\over #2}}}
\tolerance=10000
\input phyzzx

\def\unit{\hbox to 3.3pt{\hskip1.3pt \vrule height 7pt width .4pt
\hskip.7pt
\vrule height 7.85pt width .4pt \kern-2.4pt
\hrulefill \kern-3pt
\raise 4pt\hbox{\char'40}}}

\def\half{{\textstyle {1 \over 2}}}

\REF\PKTb{P.K. Townsend, {\it P-Brane Democracy}, hep-th/9507048,
to appear in the
proceedings of the March 1995 PASCOS/Johns Hopkins conference.}
\REF\pol{ J. Polchinski, Phys. Rev. Lett. {\bf 75} (1995) 184.}
\REF\polb{J. Dai, R.G. Leigh and J. Polchinski, Mod. Phys. Lett.
{\bf A4} (1989) 2073.}
\REF\ggp{G. Gibbons, M.B. Green and M.J. Perry, {\it Instantons
and Seven-Branes in
Type IIB Superstring Theory}, hep-th/9511080.}
\REF\SCS{B.R. Greene, A. Shapere, C. Vafa and S.T. Yau, Nucl.
Phys. {\bf B337} (1990) 1.}
\REF\duff{M.J. Duff and P. van Nieuwenhuizen, Phys. Lett. {\bf 94B}
(1980) 179.}
\REF\ant{A. Aurilia, H. Nicolai and P.K. Townsend, Nucl. Phys.
{\bf B176} (1980) 509.}
\REF\rom{L. Romans, Phys. Lett. {\bf 169B} (1986) 374.}
\REF\CGO{J.L. Carr, S.J. Gates, Jr. and R.N. Oerter, Phys. Lett.
{\bf 189B} (1987) 68.}
\REF\pols{ J. Polchinski and A. Strominger, {\it New Vacua for
Type II String Theory}, hep-th/9510227.}
\REF\Teit{M. Henneaux and C. Teitelboim, Phys. Lett. {\bf 143B}
(1984) 415.}
\REF\PW{E. Witten and J. Polchinski, Nucl. Phys. {\bf B460} (1996) 525.}
\REF\DHS{M. Dine, P. Huet and N. Seiberg, Nucl. Phys. {\bf B322}
(1989) 301.}
\REF\BHO{E. Bergshoeff, C.M. Hull and T. Ort\'\i n, Nucl. Phys.
{\bf B451} (1995) 547.}
\REF\SSM{J. Scherk and J.H. Schwarz, Phys. Lett. {\bf 82B} (1979)  
60.}
\REF\HT{C.M. Hull and P.K. Townsend, Nucl. Phys. {\bf B438} 
(1995) 109.}
\REF\BGRMV{S. Belluci, S.J. Gates, B. Radak, P. Majumdar and 
S. Vashakidze, Mod. Phys. Lett. {\bf A21} (1989) 1985.}
\REF\LPSS{H. L\"u, C.N. Pope, E. Sezgin and K.S. Stelle,
{\it Stainless $p$-branes}, hep-th/9508042.}
\REF\BBO{E. Bergshoeff, H.J. Boonstra and T. Ort\'\i n,
{\it $S$-duality and dyonic p-brane solutions in type II string theory},
Phys. Rev. {\bf D} {\sl in press}, hep-th/9508091.}
\REF\HS{G.T. Horowitz and A. Strominger, Nucl. Phys. {\bf B360}
(1991) 197.}
\REF\PKT{P.K. Townsend, Phys. Lett. {\bf 350B} (1995) 184.}
\REF\NEW{M.B.~Green, C.M.~Hull and P.K.~Townsend,
{\it D-Brane Wess-Zumino Actions, $T$--duality and the cosmological
constant}, {\tt hep-th/9604119}.}
\REF\BST{E. Bergshoeff, E. Sezgin and P.K. Townsend, Phys. Lett.
{\bf 189B} (1987) 75; Ann. Phys. (N.Y.) {\bf 185} (1988) 330.}
\REF\EW{E. Witten, Nucl. Phys. {\bf B443} (1995) 85.}
\REF\JHS{J.H. Schwarz, {\it The Power of M-Theory}, hep-th/9510086.}
\REF\HW{P. Ho{\v r}ava and E. Witten, Nucl. Phys. {\bf B460} (1996) 506.}
\REF\Schwarz{J.H. Schwarz, Nucl. Phys. {\bf B226} (1983) 269.}

\REF\BJO{E. Bergshoeff, B. Janssen and T. Ort\'\i n,
Class. Quantum Grav. {\bf 12} (1995) 1.}


\Pubnum{ \vbox{ \hbox{R/95/55} \hbox{hep-th/9601150} \hbox{UG-15/95} } }
\pubtype{}
\date{January 1996/revised March 1996}

\titlepage

\title {\bf Duality of Type II 7-branes and 8-branes\foot{This paper
supercedes  {\it The IIA super Eightbrane}, hep-th/9511079.}}

\author{E. Bergshoeff and M. de Roo}
\address{Institute for Theoretical Physics,
\break
University of Groningen, Nijenborgh 4,
\break
9747 AG Groningen,
\break
The Netherlands}
\andauthor{M.B. Green, G. Papadopoulos and P.K. Townsend}
\address{DAMTP, University of Cambridge,
\break
Silver St., Cambridge CB3 9EW, U.K.}
\vfill\eject

\abstract{
We present a version of ten-dimensional IIA supergravity containing
a 9-form potential for which the field equations are equivalent to
those of the standard, massless, IIA theory for vanishing 10-form 
field strength, $F_{10}$, and to those of the `massive' IIA theory 
for non-vanishing $F_{10}$. We present a multi 8-brane solution of 
these equations that generalizes the 8-brane of Polchinski and Witten. 
We show that this solution is T-dual to a new multi 7-brane solution 
of $S^1$ compactified IIB supergravity, and that the latter is
T-dual to the IIA 6-brane. When combined with the $Sl(2;\Z)$ U-duality
of the type IIB superstring, the T-duality between type II 7-branes
and 8-branes implies a quantization of the cosmological constant of 
type IIA superstring theory. These results are made possible by the  
construction of a new {\it massive} N=2 D=9 supergravity theory. We 
also discuss the 11-dimensional interpretation of these type II 
p-branes.}

\endpage


\chapter{Introduction}

Recent advances in our understanding of non-perturbative superstring  
theory
have led to the establishment of many connections between hitherto  
unrelated
superstring theories. Many of these connections involve p-brane  
solutions of
the respective supergravity theories that couple to the (p+1)-form  
potentials
in the Ramond-Ramond (RR) sector. Most  of these RR p-branes, and all  
of the
IIA ones, are singular as solutions of ten-dimensional (D=10)  
supergravity, so
their status in superstring theory was unclear until recently (see  
[\PKTb] for
a recent review). It now appears [\pol] that the RR p-branes of type  
II
supergravity theories have their place in type II superstring theory  
as
`Dirichlet-branes', or `D-branes' [\polb]. These include the p-branes  
for
$p=0,2,4,6$ in the type IIA case and the p-branes for $p=1,3,5$ in  
the type IIB
case. However, they also include a type IIB 7-brane, and a type IIA  
8-brane and
it is possible to view the D=10 spacetime as a type IIB 9-brane  
[\pol]. Note
that since the dual of a $p$-brane in D=10 is a ($6-p$)-brane, only  
$p$-branes
with $p\le6$ have duals with $p\ge0$ for which a standard 
(Minkowski space) interpretation is
available\foot{The IIB 7-brane has a (-1)-brane dual, but the latter  
has an
interpretation as an instanton [\ggp].}, so the $p$-branes with  
$p\ge7$ have
implications that are qualitatively different from those with  
$p\le6$. This
difference is also apparent in the $p$-brane solutions of the  
effective IIA or
IIB supergravity field equations. These solutions generally involve a  
function
that is harmonic on the ($9-p$)-space `transverse' to the  
($p+1$)-dimensional
worldvolume of the $p$-brane. For $p\le6$ the transverse space has  
dimension 3
or greater so there exist harmonic functions that are constant at  
infinity, but
for $p\ge7$ the transverse space has dimension 2 or less and the  
asymptotic
properties are therefore qualitatively different. Partly for this  
reason little
attention has been given so far to the $p\ge7$ branes.

Since p-branes couple naturally to (p+1)-form potentials, one expects  
to find a
stable p-brane solution of a supergravity theory only if it includes  
a
($p+1$)-form potential. From this perspective the IIB D=10 7-brane is  
the most
straightforward of the $p\ge 7$ cases because the pseudo-scalar field  
of IIB
supergravity can be exchanged for its 8-form dual. Indeed, a type IIB  
7-brane
solution has recently been described [\ggp]; it can be viewed as a  
dimensional
`oxidation' of the `stringy cosmic string' solution of [\SCS].  
However, this
class of 7-brane solutions is specific to the {\it uncompactified}  
IIB
supergravity and is therefore not expected to be related by T-duality  
to other
type II p-branes. Here we shall present a new class of multi 7-brane  
solutions
of the $S^1$-compactified IIB supergravity. In the decompactification  
limit the
new solutions reduce to the trivial D=10 Minkowski spacetime  
solution.
Nevertheless, as we shall see, these solutions are T-dual to both the  
6-brane
and the IIA 8-brane solutions of the IIA theory.

The existence of an 8-brane solution of IIA supergravity is obscured  
by the
absence of a 9-form potential, $A_9$, in the standard IIA  
supergravity theory.
However, there is one in type IIA superstring theory [\pol] and this  
suggests
that it should be possible to introduce one into the IIA supergravity  
theory.
The 9-form potential would have a 10-form field-strength $F_{10}$.  
Assuming a
standard kinetic term of the form $F_{10}^2$, the inclusion of this  
field
does not lead to any additional degrees of freedom (per spacetime  
point) and so
is not immediately ruled out by supersymmetry considerations, but it  
allows the
introduction of a cosmological constant, as explained many years ago  
in the
context of a four-form field strength in four-dimensional field  
theories
[\duff, \ant]. As it happens, a version of type IIA supergravity  
theory with a
cosmological constant was constructed (up to quartic fermion
terms) some time ago by Romans [\rom], who called it the `massive'  
IIA
supergravity theory; the complete construction via superspace methods  
was found
subsequently [\CGO]. It has been argued that the existence of the  
massive IIA
supergravity is related to the existence of the 9-form potential of  
type IIA
superstring theory [\pol]. Here we shall confirm this suggestion by
reformulating the massive IIA supergravity through the introduction  
of a 9-form
potential\foot{Strictly speaking we do this only for the bosonic
Lagrangian, but the method guarantees the existence of a fermionic
extension to a full supergravity theory.}. 
The new theory has the advantage that its solutions  
include those of
both the massless and the massive IIA theory. We propose this new IIA
supergravity theory as the effective field theory of the type IIA  
superstring,
allowing for the 9-form potential. It has been suggested [\pol,\pols]  
that the
expectation value of the dual of this 10-form field strength should  
be
interpreted as the cosmological constant of the massive IIA  
supergravity
theory. One result of this paper is the determination of the precise  
relation
between these quantities; they are conjugate variables in a sense  
discussed
previously in the D=4 context [\Teit].

The massive IIA supergravity theory has the peculiarity that D=10  
Minkowski spacetime is {\it not} a solution of the field equations 
(and neither is the product of D=4 Minkowski spacetime with a 
Calabi-Yau space). Various Kaluza-Klein (KK) type solutions were found 
by Romans but none of them were supersymmetric, i.e. his solutions 
break all the supersymmetries. A supersymmetric multi 8-brane 
configuration was recently proposed as a solution of the Killing spinor condition in an appropriate bosonic background [\PW]. We verify that
this is a solution of the field equations of the new IIA supergravity
theory and we present a generalization of it. The solutions are  
all singular at the `centres' of the metric, i.e. the 8-brane positions,
but this is a general feature of RR p-branes.

It is known that after compactification on $S^1$ the {\it  
perturbative}
type IIA and type IIB superstrings are equivalent [\polb,\DHS], being  
related
by a $\Z_2$ T-duality transformation that takes the radius $R$ of the  
$S^1$ of
one superstring theory into a radius $1/R$, in appropriate units, of  
the other
superstring theory. It follows that the {\it same} effective
N=2 D=9 field theory should be obtained by dimensional reduction of  
either the
IIA or IIB theory in D=10, and this is in fact the case [\BHO]. If  
this $\Z_2$
T-duality is valid non-perturbatively too, then $p$-brane solutions  
of the IIA
theory must correspond to $p$-brane solutions of the IIB theory and  
vice-versa,
in the sense that there are solutions of either the IIA or the IIB
theory that reduce to the same solution of the $S^1$-compactified  
theory.
In particular the double-dimensional reduction to D=9 of a given IIA  
8-brane
should be equivalent to the direct reduction to D=9 of some IIB  
7-brane. There
is a potential difficulty in verifying this because the relevant D=9  
theory
must be a {\it massive} N=2 supergravity theory. It is not too 
difficult to see how to obtain a massive N=2 D=9 supergravity 
from the massive D=10 IIA theory but it
is not so obvious how the resulting theory may also be obtained from  
the (necessarily massless) D=10 IIB theory, although it must be possible  
if T-duality is to be valid non-perturbatively. As we shall show, it is  
possible by an application of a mechanism for obtaining a massive theory
in a lower dimension from a massless one in a higher dimension.
This mechanism is essentially that of Scherk and Schwarz [\SSM] but  
in our case supersymmetry is preserved by the reduction. This result allows 
us to map
8-brane solutions of the D=10 IIA theory into 7-brane solutions of  
the IIB
theory, and vice-versa.

These IIB 7-brane and IIA 8-brane solutions may be seen as the 
effective field theory realization of the associated D-branes of
the corresponding type II superstring theory. In this context, the
$Sl(2;\R)$ symmetry of the IIB supergravity is expected to be
replaced by an $Sl(2;\Z)$ U-duality [\HT], which amounts to an
identification of points in the space $Sl(2,\R)/U(1)$ of IIB vacua
that differ by the action of $Sl(2,\Z)$. One interesting 
consequence of this IIB duality, when combined with the T-duality
of the 7-brane and 8-brane, is a quantization of the cosmological 
constant of the $S^1$-compactified IIA superstring theory\foot{We 
thank John Schwarz for suggesting this possibility to us.}.

The organisation of this article is as follows. In section 2, we  
begin with a
review of the massive IIA supergravity, introducing some  
simplifications.
In section 3, we construct the new formulation of the bosonic sector  
of this
theory, incorporating the 9-form gauge field $A_9$, in which the  
cosmological
constant emerges as an integration constant. In section 4, we  
construct
supersymmetric multi 8-brane solutions of the massive IIA  
supergravity theory,
some of which are asymptotically flat. In section 5, we show how both  
the
massive IIA supergravity and the (massless) IIB supergravity theories  
may be
dimensionally reduced to yield a new D=9 N=2 massive supergravity  
theory. We
then use this to establish the massive Type II $T$ duality rules. In  
section 6,
we construct the most general seven brane solutions of the IIB theory  
that are
both compatible with the KK ansatz and preserve half the  
supersymmetry. We then
show that the massless $T$ duality transformations take this solution  
to the
IIA 6-brane while the massive $T$-duality transformations take it to  
the IIA
8-brane solution. In section 7 we further comment on the relation to  
type IIA
superstring theory and the quantization of the cosmological constant, 
and on the connection to D=11 `M-theory'. Finally, 
in Appendix A we give a
simplified formulation of the supersymmetry transformations of IIB
supergravity.


\chapter{The massive D=10 IIA supergravity}

The bosonic field content of the massive IIA D=10 supergravity theory  
comprises
(in our notation) the (Einstein) metric, $g^{(E)}$, the dilaton,  
$\sigma$, a
massive 2-form tensor field
$B'$ and a three-form potential $C'$. One introduces the  
field-strengths
$$
\eqalign{
G &=4dC' + 6m(B')^2\cr
H &=3dB' }
\eqn\aone
$$
where $m$ is a mass parameter. The Lagrangian for these fields is  
[\rom]
$$
\eqalign{
{\cal L} &= \sqrt{-g^{(E)}}\; \Big[ R_{(E)}  
-{1\over2}|\partial\sigma|^2 -
{1\over3}e^{-\sigma}|H|^2 -{1\over12}e^{{1\over2}\sigma}|G|^2 -
m^2e^{{3\over2}\sigma}|B'|^2
-{1\over2}m^2 e^{{5\over2}\sigma}\Big]\cr
& + {1\over 9}\varepsilon \big[ dC'dC'B' + mdC'(B')^3 + {9\over20}m^2
(B')^5\big]\ .}
\eqn\atwo
$$
The notation for forms being used here is that a $q$-form $Q$ has  
components
$Q_{M_1\dots M_q}$ given by
$$
Q= Q_{M_1\dots M_q}dx^{M_1}\wedge \dots \wedge dx^{M_q}\ .
\eqn\athree
$$
Thus, the $(1/9)\varepsilon dC'dC'B'$ term in \atwo\ is shorthand for
$$
{1\over 9}\varepsilon^{M_1\dots
M_{10}}\partial_{M_1}C'_{M_2M_3M_4}\partial_{M_5}C'_{M_6M_7M_8}B_{M_9M 
_{10}}\ .
\eqn\aathree
$$

As explained in [\rom] the massless limit is not found by simply  
setting $m=0$
in \atwo\ because the supersymmetry transformations involve terms  
containing
$m^{-1}$. Instead, one first makes the field redefinitions
$$
\eqalign{
B' &= B + {2\over m}dA\cr
C' &= \tilde C - {6\over m} AdA\ . }
\eqn\afour
$$
This redefinition introduces the gauge invariance
$$
\eqalign{
\delta A &= -m\Lambda \cr
\delta B &= 2d\Lambda\cr
\delta\tilde C &= 12 Ad\Lambda}
\eqn\afive
$$
for which the gauge-invariant field strengths are
$$
\eqalign{
F&= 2dA + mB\cr
H&= 3dB\cr
G &= 4d\tilde C + 24 BdA + 6m B^2\ .}
\eqn\asix
$$
The bosonic Lagrangian of the massive IIA theory is now
$$
\eqalign{
{\cal L} &= \sqrt{-g^{(E)}}\; \Big[ R_{(E)}  
-{1\over2}|\partial\sigma|^2 -
{1\over3}e^{-\sigma}|H|^2 -{1\over12}e^{{1\over2}\sigma}|G|^2 -
e^{{3\over2}\sigma}|F|^2
-{1\over2}m^2 e^{{5\over2}\sigma}\Big]\cr
& + {1\over 9}\varepsilon \big[ d\tilde C d\tilde C B + 6d\tilde C  
B^2 dA +
12(dA)^2B^3 +
md\tilde C B^3 + {9\over2}mB^4 dA + {9\over 20}m^2 (B)^5\big]\ ,}
\eqn\aseven
$$
and the bosonic Lagrangian of the massless IIA theory can now be  
found by
taking the $m\rightarrow 0$ limit.

The Lagrangian \aseven\ can be simplified by the further redefinition
$$
\tilde C = C-6AB\ .
\eqn\aeight
$$
The $\Lambda$-gauge transformation of the new 3-form $C$ is
$$
\delta C = -6m\Lambda B
\eqn\anine
$$
and the gauge-invariant field strengths, $F$, $H$, and $G$ are now  
given by
$$
\eqalign{
F&= 2dA + mB\cr
H&= 3dB\cr
G &=4dC + 24AdB + 6mB^2 \ .}
\eqn\aten
$$
At the same time, to make contact with string theory, it is  
convenient to
introduce the string metric
$$
g_{MN} = e^{-{1\over2}\sigma}g_{MN}^{(E)}\ .
\eqn\aeleven
$$
The bosonic Lagrangian now takes the simple form
$$
\eqalign{
{\cal L} &= \sqrt{-g}\big\{e^{-2\sigma}\; \big[ R  
+4|\partial\sigma|^2
-
{1\over3}|H|^2\big]  - |F|^2 - {1\over12}|G|^2
- {1\over2}m^2\big\}\cr
& + {1\over 9}\varepsilon\big[ dC dC B + mdCB^3  + {9\over 20}m^2  
(B)^5\big]\
.}
\eqn\atwelve
$$
Observe that the final topological term is simply a type of  
Chern-Simons (CS)
term associated with the 11-form $G^2H$. Thus, the bosonic action of  
the
massive type IIA supergravity theory can be written as
$$
\eqalign{
I = \int_{{\cal M}_{10}}\! d^{10}x\; &\sqrt{-g}\Big\{ e^{-2\sigma}\;  
\Big[ R
+4|\partial\sigma|^2 - {1\over3}|H|^2\Big]  - |F|^2 -
{1\over12}|G|^2 - {1\over2}m^2\Big\} \cr
& + {1\over 9}\int_{{\cal M}_{11}}\!  G^2 H \ ,}
\eqn\athirteen
$$
where ${\cal M}_{11}$ is an 11-manifold with boundary ${\cal  
M}_{10}$. Apart
from the cosmological constant, the $m$-dependent terms in the action  
can be
simply understood as arising from the replacement of the usual  
$m$-independent
field strengths of the massless type IIA theory by their  
$m$-dependent
generalizations \aten. Furthermore, the $m$-dependence of these field  
strengths
is completely fixed by the `Stueckelberg' gauge transformation  
$\delta
A=-m\Lambda$ of $A$, as are the remaining $\Lambda$-transformations.  
The
relation of the constant $m$ appearing in this transformation with  
the
cosmological constant cannot be understood purely within the context  
of the
bosonic Lagrangian but is, of course, fixed by supersymmetry.

Observe that the cosmological constant term in \athirteen\ is now (in  
the
string
metric) independent of the dilaton. This is typical of the RR sector  
and is
consistent with the idea that $m$ can be interpreted as the  
expectation value
of the dual of a RR 10-form field strength. This interpretation would  
have the
additional virtue of restoring the invariance under the discrete  
symmetry in
which all RR fields change sign, a symmetry that is broken by the  
terms linear
in $m$ in \atwelve. We shall now show how to reformulate the massive  
IIA theory
along these lines. As we shall see the cosmological constant is  
simply related
to, but not equal to, the expectation value of the ten-form field  
strength.


\chapter{D=10 IIA supergravity with 9-form potential}

We shall start with the bosonic Lagrangian of \atwelve. Expanding in  
powers of $m$, the associated action $I(m)$ is
$$
\eqalign{
I(m) &= I(0) + \int \! d^{10} x \;  \Big\{  
2m\sqrt{-g}\big[(dC+6AdB)\cdot B^2
-2dA\cdot
B\big] + {m\over 9}\varepsilon dCB^3\big]\cr
\qquad & -{1\over2}m^2\sqrt{-g}\Big[1+2|B|^2+6|B^2|^2\Big] +{m^2\over
20}\varepsilon
B^5\Big]\Big\} \ ,}
\eqn\none
$$
where $I(0)$ is the bosonic action of the massless IIA supergravity  
theory.
We now promote the constant $m$ to a field $M(x)$, at the same time  
introducing
a 9-form potential $A_9$ as a Lagrange multipler for the constraint  
$dM=0$.
Omitting a surface term, the Lagrange multiplier term can be  
rewritten as
$$
10\; \varepsilon dA_9 M\ .
\eqn\ntwo
$$
The $A_9$ field equation implies that $M=m$, for some constant $m$,  
so the
remaining equations are equivalent to those of the massive IIA theory
except that the constant $m$ is now arbitrary and that we now have an
additional field equation from varying $M$. This additional equation  
is
$$
{\delta I(M) \over \delta M(x)} =  -\varepsilon F_{10}
\eqn\nthree
$$
where $I(M)$ is the action \none\ but with $M$ replacing $m$, and  
$F_{10}= 10\;
dA_9$ is the 10-form field strength of $A_9$. Thus the $M$ equation  
simply
determines the new field strength $F_{10}$.  Observe that the  
expectation value
of $(\varepsilon F_{10})$ is {\it not} equal to the expectation value  
of
$\sqrt{-g}M$, as a matter of principle (although it may equal it in  
special
backgrounds), but is rather the value of the variable canonically
conjugate to it.

Note that the gauge and supersymmetry transformations of the action  
$I(M)$ no
longer vanish. However, the variations of $I(M)$ are proportional to  
$dM$ and
can therefore be cancelled by a variation of the new 9-form gauge  
potential
$A_9$. This determines the gauge and supersymmetry transformations of  
$A_9$.
The supersymmetry variation will not be needed for our purposes so we  
omit it.
The $\Lambda$-gauge transformation of $A_9$ found in this way is
$$
\delta (\varepsilon A_9) = {2\over5}\sqrt{-g} \Big[\Lambda\cdot F +  
(\Lambda
B)\cdot
G\Big] - {1\over30}\varepsilon \big( 2\Lambda dC B^2 + M\Lambda  
B^4\big)\ .
\eqn\aextra
$$
We now have a new gauge-invariant bosonic action
$$
I(M) + \int d^{10}x\; M\varepsilon F_{10}\ .
\eqn\nfour
$$
The field $M$ can now be treated as an auxiliary field that can be  
eliminated
via its field equation
$$
\sqrt{-g}M = K^{-1}(B)\Big\{ \varepsilon (F_{10} + {1\over 9}dC B^3)
+2\sqrt{-g}[(dC + 6AdB)\cdot B^2 - 2dA\cdot B]\Big\}\ ,
\eqn\nfive
$$
where
$$
K(B) =  1+2|B^2|+6|B^2|^2-{1\over10\sqrt{-g}}\varepsilon B^5\ .
\eqn\nsix
$$
Using this relation in \nfour\ we arrive at the Lagrangian
$$
{\cal L}_{new} = {\cal L}_0 + \big[\sqrt{-g}K(B)\big]^{-1} \Big\{  
\varepsilon
(F_{10} +
{1\over 9}dC B^3) +2\sqrt{-g}[(dC + 6AdB)\cdot B^2 - 2dA\cdot  
B]\Big\}^2
\eqn\nseven
$$
where ${\cal L}_0$ is the bosonic Lagrangian of the massless IIA  
theory.
Note the non-polynomial structure of the new Lagrangian in the gauge  
field $B$.
This greatly obscures the $\Lambda$-gauge invariance, which is  
ensured by the
very complicated $\Lambda$-gauge transformation of $A_9$.

The full IIA supergravity Lagrangian in the new formulation of course
requires the inclusion of the fermion terms. Although we have not
worked these out it should be clear from the above construction of
the purely bosonic sector that they can be deduced directly from those
in the `old' formulation by following the above steps. Thus, the existence
of the full IIA supergravity theory with 9-form potential is guaranteed.
Presumably, there also exists an on-shell superspace formulation of
the field equations of this new theory, which it would be of interest
to find. We leave this problem to future investigations.


\chapter{The IIA D=10 Eightbrane}

The appearance of the 9-form potential in the above reformulation of  
the
massive IIA supergravity theory suggests the existence of an  
associated 8-brane
solution.  We will find solutions of the equations of motion of  
\nseven\ of the
form
$$
\eqalign{
ds^2 &= f^2(y)\; dx^\mu dx^\nu\eta_{\mu\nu} + g^2(y)dy^2\cr
\sigma &= \sigma(y) \cr
A_9 &= A_9(y)}
\eqn\bone
$$
with all other fields vanishing, and where $\eta$ is the Minkowski  
9-metric.
Such a solution will have 9-dimensional Poincar\'e invariance and  
hence an
interpretation as an 8-brane. We shall further require of such a  
solution that
it preserve some supersymmetry, so we shall begin by considering the  
variation
of the gravitino one-form $\psi$ and the dilatino $\lambda$ in the  
presence of
configurations of the above form. The full variations of the massive  
IIA theory
can be found in [\rom] in the Einstein--frame. They depend on the  
constant
$m$. In the new theory, this constant is replaced by the function $M$  
given in \nfive. Here, however, we shall need the fermion variations 
in the {\it string-frame}. For $M=0$ these are implicit in the superspace results of [\BGRMV]. For the backgrounds considered here, for which 
all fermions vanish and $\sqrt{-g}M=\varepsilon F_{10}$,
the $M\ne 0$ string-frame fermion variations are most easily deduced
from the Einstein-frame results of [\rom]. The result is 
$$
\eqalign{
\delta_\epsilon \psi &= D\epsilon + {1\over 8}M  
e^{\sigma}\Gamma\epsilon\cr
\delta_\epsilon \lambda &= -{1\over  
2\sqrt{2}}\big(\Gamma^M\partial_M\sigma +
{5\over4}Me^{\sigma}\Big)\epsilon\ . }
\eqn\btwo
$$

For configurations of the assumed form, and further assuming that  
$\epsilon$
depends only on $y$, the equations $\delta\psi=0$ and  
$\delta\lambda=0$ become
$$
\eqalign{
0&=g^{-1}\epsilon' +{1\over 8}M e^{\sigma} \bar\Gamma_y\epsilon
\cr
0& = \big( g^{-1}f'\bar\Gamma_y
+{1\over 4}M fe^{\sigma} \big) \epsilon \cr
0&= \big(g^{-1}\sigma' +  
{5\over4}Me^{\sigma}\bar\Gamma_y\big)\epsilon \ ,}
\eqn\bthree
$$
where the prime indicates differentiation with respect to $y$ and  
$\bar\Gamma$
are the {\it constant}, orthonormal frame basis, gamma--matrices. To  
find
non-zero solutions for $\epsilon$ we are now forced to suppose that  
$\epsilon$
has a definite `chirality' in the sense that
$$
\bar\Gamma_y\epsilon =\pm \epsilon\ .
\eqn\bfour
$$
We then find that
$$
g^{-1}f' = \mp {1\over 4}Mfe^{\sigma}
\eqn\abfive
$$
and that
$$
g^{-1}\Big(e^{-\sigma}\Big)' =\pm {5\over 4}M\ .
\eqn\bfive
$$
Eliminating $M$ from these equations we deduce that $f$ is  
proportional to
$e^{\sigma/5}$. As we are free to rescale the coordinates $x^\mu$ we  
may choose
the constant of proportionality to be unity, without loss of  
generality. Thus,
$$
f= e^{{1\over 5}\sigma}\ .
\eqn\bsix
$$

We are also free to choose $g(y)$ to be any function that is  
non-singular where $f(y)$ is non-singular\foot{i.e. where 
neither $f$ nor $f^{-1}$ vanish.}. For example, the 
choice $g=f$ leads to a manifestly
conformally flat form of the 8-brane metric. A solution in this form  
was given
in [\PW]. We postpone a discussion of this solution until we have the  
general
solution, to be given below. The choice of $g$ that we shall make  
here is
$g=f^{-1}$. In this case, use of the $A_9$ field equation $M'=0$ in  
\bfive\
yields
$$
\partial_y^2\Big( e^{-{4\over5}\sigma}\Big) =0\ .
\eqn\bsixa
$$
The general solution is given in terms of a harmonic function $H(y)$,
the precise nature of which will be discussed shortly, i.e.
$$
e^{-{4\over5}\sigma}= H(y)\ .
\eqn\bseven
$$
This leads to the 8-brane configuration
$$
\eqalign{
ds^2 &= H^{-{1\over 2}}\; dx^\mu dx^\nu\eta_{\mu\nu}\  + \  
H^{1\over2}dy^2\cr
e^{-4\sigma} &= H^5 \cr
M &= \pm H^\prime \ .}
\eqn\bnine
$$
where the prime indicates differentiation with respect to $y$. We  
have verified
that this configuration is a solution of the full set of field  
equations.
The Killing spinor $\epsilon$ is given by
$$
\epsilon = H^{-{1\over8}}(y)\epsilon_0,\qquad \bar\Gamma_y\epsilon_0  
=\pm
\epsilon_0\ , \eqn\bseven
$$
where $\epsilon_0$ is a constant spinor.

It remains to determine the harmonic function $H$. Consider first the  
massive
IIA theory for which the function $M$ equals the (non-zero) constant  
$m$
appearing in the Lagrangian, which we may choose to be positive. In  
this case
$$
H=\pm m(y-y_0)
\eqn\gone
$$
for constant $y_0$, where the sign depends on the choice of  
'chirality' of
$\epsilon$. However, $H$ must be positive for real $\sigma$, so the  
spinor
$\epsilon$ must change chirality at $y=y_0$. This is 
possible because  
$\epsilon$ blows up at $y=y_0$\foot{Presumably, this 
is acceptable because the metric is also singular at 
$y=y_0$; in any case, we shall see below that this 
feature is not generic for the general 8-brane 
solution of the new IIA supergravity theory.}.
Thus, the massive IIA theory has a solution for which
$$
H= m|y-y_0|\ .
\eqn\gtwo
$$
Note that this is a continuous function of $y$ with a kink  
singularity at
$y=y_0$, at which the curvature tensor has a delta function  
singularity.

In the new IIA theory we may suppose that $M$ is only {\it locally}
constant. The form of the function $H$ in this case depends on the  
type of
point singularity that we allow. The above example suggests that we  
should
require $H$ to be a continuous function of $y$. There are solutions  
for which
$H$ is discontinuous but they have $\delta'$ type singularities of  
the
curvature tensor, and we shall not consider them. In any case, the  
restriction
to kink singularities produces physically sensible results, as we  
shall see.
An example of a solution with a single kink singularity of $H$ is
$$
H=\cases{-ay+b \qquad y<0\cr cy+b\qquad y>0}
\eqn\gthree
$$
where $a$, $b$ and $c$ are non-negative constants. We adopt this as  
the
basic single 8-brane solution. It can be interpreted as a domain wall
separating regions with different values of $M$. The regions  
$y\rightarrow
\pm\infty$ are
at infinite affine distance. The solution therefore has two  
asymptotic regions
relative to which an 8-brane charge, $Q_\pm$, may be defined as the  
value of
$M$ as $y\rightarrow \pm \infty$. For the above solution,
$$
Q_+ =c\qquad Q_-= a \ .
\eqn\gfour
$$
The constant $b$ determines the value of $\sigma$, and hence the  
value of the
string coupling constant $e^\sigma$ at the 8-brane core. In  
particular, if
$b=0$ the string coupling constant goes to infinity at the core. Note  
that the
solution \gtwo\ of the masive IIA theory is the special case for  
which $a=c$
and $b=0$.

The multi 8-brane generalization of \gthree\ with the same charges is
found by allowing kink singularities of H at n+1 ordered points
$y=y_0<y_1<y_2<\dots <y_n$. The function $H$ is
$$
H=\cases{-a(y-y_0) + \sum_{i=1}^n\mu_i (y_i-y_0) + b\qquad y<y_0 \cr
(c-\sum_{i=1}^n\mu_i) |y-y_0| + \sum_{i=1}^n \mu_i |y-y_i| +b \qquad  
y>y_0}
\eqn\gfive
$$
where $\mu_i$ are positive constants and $a$, $b$, $c$
are non-negative constants.

The asymptotically left-flat or right-flat solutions are those for  
which
$Q_-=0$ or $Q_+=0$, respectively. The asymptotically flat solutions  
are those
which are both asymptotically left-flat and right-flat. An example of  
an
asymptotically flat three 8-brane solution is given by $H=
\mu^2\big||y-y_0|-|y-y_1|\big| + \gamma^2$, where $\mu$ and $\gamma$  
are
arbitrary constants.

If we now introduce a new variable $w(y)$ such that
$$
{ dw\over dy} = H^{{1\over2}}\ ,
\eqn\camone
$$
then the above 8-brane solution becomes
$$
\eqalign{
ds^2 &= Z^{-{1\over3}}(w)\big[ dx\cdot dx + dw^2\big]\cr
e^{-\sigma} &= Z^{5\over 6}(w) \ ,}
\eqn\camtwo
$$
where $Z(w)$ is a harmonic function of $w$, related to $H(y)$ by
$$
Z(w) = H^{3\over2}\big(y(w)\big)\ .
\eqn\camthree
$$
The 8-brane solution of [\PW] is of this form. For example, the  
single 8-brane of that reference
corresponds to the special choice of $H$ in \gthree\ with  
$b=0$. This
can be seen from the fact that the conformal factor of the solution  
of [\PW]
blows up at the position of the 8-brane, while this is true of the  
above
8-brane solution only if $b=0$.

If the above 8-brane solutions are to be considered as a field theory
realization of the Dirichlet 8-brane of type IIA superstring theory  
then one
would expect them to be related by T-duality to both 7-brane and  
9-brane
solutions of IIB
supergravity. Consider first the Dirichlet 9-brane; its field theory
realization is just D=10 Minkowski spacetime or, in the context of  
the $S^1$
compactified theory required for T-duality considerations, the  
product of $S^1$
with D=9
Minkowski spacetime. This spacetime can also be regarded as an  
8-brane
solution.
T-duality requires that the same solution result from direct (as  
against
double) dimensional reduction of the 8-brane solution found above.  
This is
indeed the case because compatibility with the KK reduction requires  
us to
choose the harmonic function $H$ to be constant, in which case $M=0$.  
The
dimensionally reduced solution is then precisely the product of $S^1$  
with D=9
Minkowski spacetime.

Thus, the IIB 8-brane can be regarded as a T-dual of the IIB 9-brane.
The more difficult task of determining the IIB 7-brane solutions to  
which the
IIA 8-brane solutions are T-dual is what will occupy us for most of  
the
remainder of this paper; it involves the construction of a new  
massive D=9
supergravity, to which we now turn our attention.


\chapter{Massive D=9 N=2 supergravity}

The standard dimensional reduction to D=9 of either the massless IIA
supergravity theory or the IIB supergravity theory yields the  
massless N=2 D=9
supergravity theory [\BHO] (see also [\LPSS]). Here we shall  
construct a
massive N=2 D=9 supergravity theory. We shall do this in two ways.  
The first
involves the massive IIA supergravity theory. At first sight it might  
seem
that this theory cannot be dimensionally reduced to D=9 because the  
product of
D=9 Minkowski space with $S^1$ is not a solution of the field  
equations.
However, all we need is a solution with an abelian isometry and the  
massive IIA
8-brane is such a solution. This allows us to reduce the massive D=10  
IIA
theory to D=9\foot{An alternative, equivalent, procedure would be to  
make use
of the new $A_9$ formulation of IIA supergravity to reduce to D=9 in  
the
standard
way; the resulting D=9 theory has an 8-form potential which can be  
traded for a
cosmological constant.}. We shall then show that exactly the same  
theory can be
found by a Scherk-Schwarz dimensional reduction of the IIB  
supergravity theory.

We begin by dimensionally reducing the massive IIA supergravity  
theory. Since
we ultimately wish to make contact with the IIB theory via T-duality,  
it is
convenient to use the conventions of [\BHO], where the massless  
T-duality rules
are given. Thus, the first step is to rewrite the results of section  
2 in the
notation of [\BHO]. The field content in D=10 is given by
$$
\bigl \{{\hat g}_{\hat\mu\hat\nu}, {\hat C}_{\hat\mu\hat\nu\hat\rho},
{\hat B}^{(1)}_{\hat\mu\hat\nu}, {\hat A}^{(1)}_{\hat\mu}, \hat\phi
\bigr\}
\eqn\ninea
$$
where the fields $\hat C$ and ${\hat A}^{(1)}$ are the R-R sector  
fields. We
refer to [\BHO] for details of the notation, but we remark 
here that  
{\it in this section only the metric signature is `mostly minus'} 
and that the hats
indicate D=10 variables; the D=9 variables resulting from the  
dimensional
reduction will be without hats. Our starting point is the following
(string-frame) action, obtained by translating \athirteen\ 
into the conventions of [\BHO]:
$$
\eqalign{
I_{\rm IIA} = &{1\over 2}\int_{{\cal M}_{10}} d^{10} x \sqrt {-\hat  
g}
\biggl \{ e^{-2\hat\phi}\bigl [ -\hat R + 4|d\hat\phi|^2 - {3\over 4}
|{\hat H}^{(1)}|^2\bigr ]\cr
&+{1\over 4}|{\hat F}^{(1)}_{m}|^2 + {3\over 4}|{\hat G}_{m}|^2
+ {1\over 2}m^2\biggr \} +{1\over 64} \int_{{\cal M}_{11}}{\hat  
G}^2_m{\hat
H}^{(1)}\ .}
\eqn\nineb
$$
Apart from the cosmological term, all $m$ dependent terms occur via  
the
field-strength tensors of the R-R fields. As explained in section 2,  
the
$m$-dependent terms within these curvature tensors are determined by  
the
Stueckelberg type symmetries, which now read:
$$
\eqalign{
\delta{\hat B}^{(1)} &= d{\hat \eta}^{(1)}\cr
\delta {\hat A}^{(1)} &= -{m\over 2}{\hat \eta}^{(1)}\cr
\delta \hat C &=-m{\hat \eta}^{(1)}{\hat B}^{(1)}\ .}
\eqn\ninec
$$
The $m$-dependence of the corresponding R-R curvatures is given by
$$
\eqalign{
{\hat F}^{(1)}_m &= {\hat F}^{(1)}_{m=0} + m{\hat B}^{(1)}\cr
{\hat G}_m &= {\hat G}_{m=0} + {m\over 2}({\hat B}^{(1)})^2\ .}
\eqn\nined
$$
The $m=0$ part of the curvatures in the conventions now being used  
may be found
in [\BHO].

The field content of the massive $D=9$ Type II theory is given by
$$
\bigl \{g_{\mu\nu}, C_{\mu\nu\rho},
B^{(i)}_{\mu\nu}, {A}^{(i)}_{\mu}, \phi, k, \ell
\bigr\}\ .
\eqn\ninee
$$
The R-R sector fields are $C, B^{(2)}, A^{(1)}$ and $\ell$.
The action can be obtained by straightforward dimensional reduction
of the ten-dimensional theory and is given by
$$
\eqalign{
I = &{1\over 2}\int_{{\cal M}_9} d^9x \sqrt g
\biggl\{ e^{-2\phi}\bigl [ -R + 4|d\phi|^2 - {3\over 4}|H^{(1)}|^2\cr
& -|d{\rm log} k|^2 + {1\over 4}k^2|F^{(2)}|^2 + {1\over 4}k^{-2}
|F(B)|^2\bigr ] + {1\over 2} m^2 k\cr
&- {1\over 2}k^{-1}|d\ell -mB|^2 +{1\over 4}k|F_m^{(1)}|^2
+{3\over 4}k|G_m|^2 - {3\over 4}k^{-1}|H_m^{(2)}|^2\biggr\}\cr
&-{1\over 64}\int_{{\cal M}_{10}} G_m^2F(B) + 4G_mH^{(1)}H^{(2)}_m \  
.}
\eqn\ninef
$$
The $m$-dependent factors in the curvature tensors are determined by
the following $D=9$ Stueckelberg type symmetries (which follow
straightforwardly from the $D=10$ rules)
$$
\eqalign{
\delta B &= d\Lambda\cr
\delta A^{(1)} &= -{m\over 2}\eta^{(1)} -m\Lambda A^{(2)}\cr
\delta B^{(1)} &= d\eta^{(1)} - A^{(2)}d\Lambda\cr
\delta B^{(2)} &= A^{(1)}d\Lambda + m\Lambda B^{(1)} +
{m\over 2}\eta^{(1)}B\cr
\delta C &= -m\eta^{(1)}\bigl ( B^{(1)} + A^{(2)}B\bigr )\cr
\delta \ell &= m\Lambda \ .}
\eqn\nineg
$$
These Stueckelberg symmetries lead to the following (unique)
modified curvatures for the R-R fields:
$$
\eqalign{
F^{(1)}_m &= F^{(1)}_{m=0} + \ell F^{(2)}_{m=0} + m\bigl (B^{(1)}
-A^{(2)}B\bigr )\cr
G_m &= G_{m=0} + {1\over 2}m(B^{(1)})^2 - mB^{(1)}A^{(2)}B\cr
H^{(2)}_m &= H^{(2)}_{m=0} -\ell H^{(1)}_{m=0} -mBB^{(1)}\, .}
\eqn\nineh
$$
The expressions for the $m=0$ curvatures may again be found in  
[\BHO].

We now turn to the (massless) $D=10$ Type IIB theory. Its field  
content is given by
$$
\bigl \{{\hat \j}_{\hat\mu\hat\nu},{\hat {\cal  
B}}^{(i)}_{\hat\mu\hat\nu},
\hat\ell, \hat\varphi, {\hat  
D}_{\hat\mu\hat\nu\hat\rho\hat\sigma}^{(+)}
\bigr\}\, ,\hskip 1.5truecm  i=1,2.
\eqn\ninei
$$
The R-R sector fields are ${\hat {\cal B}}^{(2)},
\hat D^{(+)}$ and $\hat\ell$.
The action is given by\foot{Strictly speaking, there is no
action for the $D=10$ Type IIB theory.
However, when properly used, the given action leads to a
well-defined action in $D=9$. For more details about this point, see
e.g. [\BBO].}
$$
\eqalign{
I_{IIB} = {1\over 2}\int_{{\cal M}_{10}}d^{10}x &{\sqrt {-\j}}
\biggl\{e^{-2\hat\varphi}\bigl [ -\hat R + 4|d\hat\varphi|^2
-{3\over 4}|{\hat {\cal H}}^{(1)}|^2\bigr ]\cr
&-{1\over 2}|d\hat\ell|^2 -{3\over 4}|{\hat {\cal H}}^{(2)} -  
\hat\ell
{\hat {\cal H}}^{(1)}|^2 - {5\over 6}|{\hat F}(D)|^2\biggr\}\cr
&-{1\over 96}\int_{{\cal M}_{11}}\epsilon^{ij}{\hat F}(D){\hat
{\cal H}}^{(i)}{\hat {\cal H}}^{(j)}\ .}
\eqn\ninej
$$

The question now is whether, after dimensional reduction to $D=9$,  
the
massless $D=10$ Type IIB theory can be mapped onto the massive $D=9$
Type II theory found above. The standard reduction is given in [\BHO]  
and leads
to the massless theory in nine dimensions. Since one cannot add a  
cosmological
constant to the $D=10$ Type IIB theory we have to change something in  
the
standard reduction. Our guiding point will be the $D=9$
Stueckelberg symmetries \nineg. Once we can reproduce these,
the action, with the exception of the cosmological term, follows by  
symmetry.

We observe that from the IIA point of view the Stueckelberg
$\Lambda$-transformation  is just the $\underline x$-component of the
$D=10$ Stueckelberg symmetry. From the IIB point of view it should  
come
from a general coordinate transformation in the ${\underline x}$  
direction
since we know that ${\hat \xi}^{\underline x} = \Lambda$ for $m=0$
[\BHO]. In order to reproduce the  Stueckelberg  
$\Lambda$-transformations we
should therefore introduce an extra $\underline x$ dependence in some  
of the
fields of the $D=10$ IIB theory. The only $D=9$ fields that have an
$m$-dependent $\Lambda$-transformation are $\ell$ and $B^{(2)},  
A^{(1)}$.
We find that these R-R fields can be given the correct $\Lambda$
transformation provided we introduce the following additional  
dependence linear
in $\underline x$:
$$
\eqalign{
\hat \ell &= \ell +m{\underline x}\cr
{\hat {\cal B}}_{\mu\nu}^{(2)} &=
B_{\mu\nu}^{(2)} - B_{[\mu}A_{\nu]}^{(1)}
+m{\underline x}\bigl ( B_{\mu\nu}^{(1)} +  
B_{[\mu}A_{\nu]}^{(2)}\bigr )\cr
{\hat {\cal B}}^{(2)}_{{\underline x}\mu} &= - A_\mu^{(1)} +
m{\underline x}A_\mu^{(2)}\ .}
\eqn\ninek
$$
Note that the $x$-dependence in $\hat \ell$, which was introduced to  
reproduce
the correct $D=9$ Stueckelberg $\Lambda$-transformation, at the same  
time
leads, via the kinetic term of $\hat \ell$, to the desired  
cosmological
constant in $D=9$! This establishes a relation between the  
cosmological
constant
and the Stueckelberg symmetries. Note also that, although the
IIB R-R fields $\hat \ell$ and ${\hat {\cal B}}^{(2)}$ depend on
$\underline x$, all the $\underline x$-dependence drops out in the  
$D=10$
IIB action. This can be seen by rewriting the ansatz for ${\hat {\cal
B}}^{(2)}$ in the following equivalent form:
$$
{\hat {\cal B}}^{(2)}_{\hat \mu\hat\nu}
= {\hat {\cal B}}^{(2)}_{\hat\mu\hat\nu, m=0} + m{\underline x}
{\hat {\cal B}}^{(1)}_{\hat\mu\hat\nu, m=0\ .}
\eqn\ninel
$$

Finally, we still have to reproduce the correct $\eta^{(1)}$  
Stueckelberg
symmetries. For $m=0$ this symmetry is related to the following
Type IIB gauge symmetry:
$$
\eqalign{
\delta {\hat {\cal B}}^{(i)} &= d{\hat \Sigma}^{(i)}\cr
\delta \hat D &= {3\over 4}d{\hat \Sigma}^{(2)}{\hat {\cal B}}^{(1)}
-{3\over 4}d{\hat \Sigma}^{(1)}{\hat {\cal B}}^{(2)}\ ,}
\eqn\ninem
$$
with ${\hat {\Sigma}}^{(i)} = \eta_\mu^{(i)}$. It turns out that the  
following
$\underline x$-dependence in ${\hat {\Sigma}}^{(2)}$ reproduces the
correct $\eta^{(1)}$ Stueckelberg symmetry given in \nineg :
$$
\hat \Sigma_\mu^{(2)} = \eta_\mu^{(2)} + m{\underline  
x}\eta_\mu^{(1)}\, .
\eqn\ninen
$$
This equation also follows from the requirement that the ansatz for
${\hat {\cal B}}^{(2)}_{\mu\nu}$ be consistent with the $m=0$ rule
$\delta B^{(1)}= d\eta^{(1)}$.

We have therefore recovered by non-trivial dimensional reduction of  
IIB
supergravity the massive N=2 D=9 supergravity found earlier from  
reduction of
the massive IIA theory. It is of interest to see how this mechanism  
is related
to the Scherk-Schwarz (SS) mechanism [\SSM]. The essential ingredient  
in their
method was a global $U(1)$ symmetry in the higher dimension.
Let $Q$ be the anti-hermitian generator of this $U(1)$ symmetry and  
let
$\partial$ denote differentiation with respect to the KK 
coordinate. Then the SS
mechanism can be summarised by the equation $\partial = mQ$. In our  
case the
relevant $U(1)$ group acts on $\hat\ell$ (which is periodically  
identified) by
a shift, so we should require $\partial \hat\ell =m$. 
The solution is  
$\hat\ell =\ell + m\underline {x}$, as above. The $U(1)$  
transformation of the
field strength three-forms $\hat{\cal H}$ must be such that the  
action \ninef\
is invariant, which determines the action of $Q$ on these fields.  
Setting
$\partial =mQ$ then yields a dependence of these field strengths on
${\underline x}$ that is consistent with the ${\underline  
x}$-dependence
\ninel\ of the two-form potentials.

Thus, the dimensional reduction used above is essentially an  
application of the
SS method. However, the implications are rather different in the  
present
context. For example, in the reduction of D=4 N=1 supergravity to D=3  
using the
global chiral $U(1)$ symmetry [\SSM], the SS mechanism  generates  
masses for
the fermions but no scalar potential, thereby breaking supersymmetry.  
In
contrast, in our case a cosmological constant is also generated and  
the full
supersymmetry of the action is preserved by the reduction. The reason  
that
supersymmetry is preserved can be traced to the fact that the  
fermions can be
redefined in such a way that they are $U(1)$-invariant. The specific
redefinition required for this is given in Appendix A for the  
Einstein-frame
fields but since the dilaton is $U(1)$ invariant this result holds  
also for the
string-frame fields. Alternatively, one can note that in the example  
given in
[\SSM] the chiral $U(1)$ acts on the supersymmetry parameter and this  
triggers
the Higgs mechanism.

Having established that both the $D=10$ massive type IIA and massless
type IIB theory map onto the same $D=9$ massive Type II supergravity  
theory,
it is straightforward to determine the massive type II $T$ duality
rules. We consider here only the map from massive IIA to massless  
IIB.
The $m=0$ rules are given in [\BHO]. We give here only the rules that  
receive
an $m$-dependent correction. These are the following:
$$
\eqalign{
\hat \ell &= {\hat A}^{(1)}_{{\underline x}} + m{\underline x}\cr
{\hat {\cal B}}_{\mu\nu}^{(2)} &=
{3\over 2}{\hat C}_{\mu\nu{\underline x}} - 2 {\hat A}^{(1)}_{[\mu}
{\hat B}^{(1)}_{\nu]{\underline x}} +
2{\hat g}_{{\underline x}[\mu}{\hat B}^{(1)}_{\nu]x}
{\hat A}_{{\underline x}}^{(1)}/g_{\underline {xx}}\cr
&+m{\underline x}\biggl ({\hat B}^{(1)}_{\mu\nu} +
2{\hat g}_{{\underline x}[\mu}{\hat B}^{(1)}_{\nu] {\underline x}}
g_{\underline {xx}}\biggr )\cr
{\hat {\cal B}}^{(2)}_{{\underline x}\mu} &= -{\hat A}_\mu^{(1)}
+ {\hat A}^{(1)}_{\underline x}
{\hat g}_{{\underline x}\mu}/{\hat g}_{\underline {xx}}\cr
& +m{\underline x}{\hat g}_{{\underline x}\mu}/{\hat g}_{\underline  
{xx}}\ . }
\eqn\tena
$$


\chapter{The circularly symmetric IIB 7-brane}

The massive $T$--duality rules derived in the previous section are  
expected to
relate IIA 8-brane solutions to IIB 7-brane solutions. Compatibility  
of the
latter with the KK ansatz implies that the 7-brane solution must have  
circular
symmetry in the transverse directions. We therefore begin with a  
construction
of the general IIB 7-brane solution of this type that also preserves  
half the
supersymmetry.

The most general static circularly symmetric 7-brane metric is
$$
ds^2 = f^2(r) d{\tilde x}\cdot d{\tilde x} + a^2(r) \big (
d\chi + \omega(r) dr \big )^2 + b^2(r)dr^2\, ,
\eqn\iione
$$
where $\{\tilde x\}$ are the coordinates of 8-dimensional Minkowski  
spacetime
(the 7-brane worldvolume), the $\chi$--coordinate is along the  
$U(1)$ Killing
vector field and $r$ is a radial coordinate. We are free to choose  
the function
$b$, and we shall choose it such that $b=a$.  Next, we change  
coordinate from
$\chi$ to
$$
{\underline x}= \chi + \kappa(r)\,  ,
\eqn\iitwo
$$
where the function $\kappa$ is such that
$$
\omega = {d\kappa\over dr}\, .
\eqn\iitwoa
$$
Note that since $\chi$ was an angular coordinate, so also is ${\underline x}$.
We may choose the identification such that
$$
{\underline x}\sim {\underline x} +1\ .
\eqn\iitwob
$$

The metric now reads
$$
ds^2 = f^2(r) d{\tilde x}\cdot d{\tilde x} + a^2(r)\big[ d{\underline  
x}^2 +
dr^2\big]\, .
\eqn\iithree
$$
To find supersymmetric solutions, we assume that
$$
\eqalign{
\hat \ell &= \ell(r) + \tilde m {\underline x}\, ,\cr
\hat \varphi &=  \hat\varphi (r) }
\eqn\iifour
$$
where $\tilde m$ is piecewise constant, and we set the rest of the 
fields equal to zero. 

Next, we substitute this ansatz into the (string frame) Killing  
spinor equations
$$
\eqalign{
\delta_\epsilon \psi &\equiv D\epsilon + {1\over 8} ie^{\hat\varphi}
\big (\Gamma^M\partial_M\hat\ell\big ) \Gamma\epsilon = 0\, ,\cr
\delta_\epsilon\lambda &\equiv {1\over 4} \bigg
(\Gamma^M\partial_M\hat\varphi +  
ie^{\hat\varphi}\Gamma^M\partial_M\hat\ell
\bigg )\epsilon = 0\, ,}
\eqn\iifive
$$
and assume that $\epsilon=\epsilon(r)$ where
$$
\bar\Gamma_{\underline x}\epsilon(r) = \pm i\bar\Gamma_r\epsilon(r)\ . 
\eqn\iiifive
$$
We thereby deduce that
$$
\epsilon^\prime \pm {1\over8} e^{\hat\varphi}\tilde m \epsilon =0
\eqn\aiifive
$$
and
$$
\eqalign{
f^\prime \pm {1\over 4} e^{\hat\varphi} f \tilde m  &= 0\, ,\cr
\ell^\prime &= 0\, ,\cr
a^{-1}a^\prime \mp{1\over 4} \tilde me^{\hat\varphi} &= 0\, ,\cr
\hat\varphi^\prime \pm e^{\hat\varphi} \tilde m  &= 0\, ,}
\eqn\iisix
$$
where the prime indicates differentiation with respect to $r$.

Using the last two equations in \iisix\ ,  we have that
$$
\partial^2_r \big ( e^{-\hat\varphi}\big ) = 0\, .
\eqn\iiseven
$$
Thus we can set
$$
e^{-\hat\varphi}  = H(r)\, ,
\eqn\iieight
$$
where $H$ is a harmonic function of $r$, of the type described in  
section 4. The last of equations \iisix\ now yields
$$
\tilde m =\pm H^\prime\ ,
\eqn\iieighta
$$
while the remainder of equations \iisix\ yields the full 
7-brane solution in terms of $H$ and three constants of 
integration, which can be removed by rescaling the coordinates 
and shifting $\hat\ell$. This solution is
$$
\eqalign{
ds^2 &= H^{-{1\over2}}(r) d{\tilde x}\cdot d{\tilde x} +
H^{1\over2}(r)\big[ d{\underline x}^2 + dr^2\big] \, ,\cr
e^{-\hat\varphi} &= H(r)\cr
\hat\ell &= \pm H'(r) {\underline x}\, .}
\eqn\iiten
$$
The Killing spinor corresponding to this solution is given by
$$
\epsilon = H^{-{1\over8}} \epsilon_0\, ,\qquad \bar\Gamma_{\underline  
x }\epsilon_0 =
\pm i \bar\Gamma_r\epsilon_0\, .
\eqn\iieleven
$$

We suggest that this 7-brane solution of IIB 
supergravity is the field theory
realization of the Dirichlet 7-brane of type IIB superstring theory.  
As a
check on this interpretation we shall now verify that it is T-dual to  
the IIA
6-brane solution of [\HS]. To this end we take $\{\tilde x\} = (v^m,  
u)$, where
$v^m$ are coordinates for 7-dimensional Minkowski spacetime (the  
6-brane
worldvolume), and also take the ignorable coordinate $u$ to be an  
angular
coordinate. We can then apply (massless) T-duality rules of [\BHO],  
in the $u$
direction. This leads to the following solution of IIA supergravity:
$$
\eqalign{
ds^2 &= H^{-{1\over2}}(r)\; dv\cdot dv \  + \ H^{1\over2}(r)
\big[du^2 + d{\underline x}^2 + dr^2 \big]\cr
e^{-\hat\phi} &= H^{3\over4}(r)\cr
\hat A^{(1)} &= \pm H^\prime (r){\underline x} du\ .  }
\eqn\othertwo
$$
This is precisely the IIA 6-brane solution in the form given in  
[\PKT], except
that in the general 6-brane solution the harmonic function $H$  
depends on all
three `transverse' variables $(u,{\underline x},r)$. Thus, this is  
the form of
the 6-brane compatible with a KK reduction to D=8. This is an  
encouraging sign
that the 7-brane will also be T-dual to a IIA 8-brane solution, since  
one
expects the 6-brane and 8-brane to be equivalent on reduction to D=8.

In order to show that this is indeed the case we need to establish  
the
T-duality of the 7-brane to the 8-brane. We shall now show that the  
massive
$T$--duality rules that we have given in section 5 relate the IIA  
eight--brane
of section 4
to the IIB seven--brane given in \iiten. Although the general massive  
Type II
$T$ duality rules are complicated they become very simple for the  
special
solutions considered here. Since
$$
{\hat g}_{{\underline x}\mu} = \hat C = {\hat B}^{(1)} =
{\hat A}^{(1)} = 0\ ,
\eqn\tenb
$$
for our solutions, the massive Type II $T$ duality rules
are
$$
\eqalign{
{\hat \j}_{\mu\nu} &= {\hat g}_{\mu\nu}\cr
{\hat \j}_{\underline {xx}} &= 1/{\hat g}_{\underline {xx}}\cr
{\hat \ell } &= m{\underline x}\cr
{\hat \varphi} &= {\hat \phi} - {1\over 2}{\rm log}
(-{\hat g}_{\underline {xx}}) \ .}
\eqn\tenc
$$

To show that under the massive $T$ duality rules the IIA eight brane  
solution
of section 4 is $T$ dual to the IIB seven brane solution, \iiten,
we first make the change of notation $\{x^\mu\} = (\tilde x,  
{\underline x})$
and $y=r$. We then wrap the eight brane in a compactifying direction,  
which we
can choose to be ${\underline x}$. The 8-brane solution is then
as follows:
$$
\eqalign{
ds^2 &=  H^{-{1\over2}}(r) d{\tilde x}\cdot d{\tilde x} +
H^{-{1\over 2}}(r)d{\underline x}^2 + H^{1\over2}(r) dr^2 ,\cr
e^{-4\sigma} &= H^5(r) \cr
M &= \pm H^\prime(r)\, ,}
\eqn\iitwelve
$$
It is now straightforward to show that the T--duality rules, 
\tenc, applied to the ${\underline x}$ direction, take
the IIA eight brane solution to the IIB seven brane solution.
The equivalence of eight and seven branes 
is a non-trivial check of our massive $T$--duality rules.


\chapter{Superstrings and Supermembranes}

The metric and dilaton ($\sigma$) of the RR p-brane solutions of D=10  
IIA or
IIB supergravity were shown in [\PKT], for $p\le 6$, to be  
expressible in the
form
$$
\eqalign{
ds^2 &= H^{-{1\over2}} d^2s_{p+1} + H^{1\over2} d{\bf y}\cdot d{\bf  
y}\cr
e^{4\sigma} &= H^{3-p} \ ,}
\eqn\sevenone
$$
where $d^2s_{p+1}$ is the Minkowski (p+1)-metric, 
$d{\bf y}\cdot d{\bf y}$ is the Euclidean metric on the  
`transverse'
space $\R^{9-p}$, and $H$ is a harmonic function on this space, 
apart from point
singularities. Here we have found new IIB 7-brane solutions that are  
also of
this form and we have shown that they are related by T-duality both  
to the IIA
KK 6-brane and to the IIA 8-brane, of which we have also given the  
general
solution preserving half the supersymmetry. This 8-brane solution 
can also be put in the above form. One advantage of the  
this form of the solutions is that the T-duality between 
the RR p-branes and the RR (p+1)-branes, after
compactification on $S^1$, is an almost immediate consequence of the  
T-duality
rules of [\BHO], at least for $p\le 6$. The relationship between the  
7-brane
and the 8-brane solutions is more subtle, as we have seen, because it  
involves
the comparison in D=9 via a previously unknown {\it massive} N=2 D=9
supergravity theory.

So far, the context of our discussion has been that of supergravity 
rather than superstring theory. A new feature of the IIB superstring 
theory is its conjectured $Sl(2;\Z)$ U-duality which requires, 
in particular, that the pseudoscalar $\hat\ell$ be periodically 
identified, i.e. that it take values in $S^1$. 
Without loss of generality 
we can suppose that the identification is such that
$$
\hat\ell \sim \hat\ell +1
\eqn\iifoura
$$
Returning now to the ansatz \iifour, we note that since ${\underline x}\sim {\underline x}+1$, the consistency of this ansatz requires $\tilde m$ to be 
an integer. Of course, since $\tilde m$ is not dimensionless, this result 
holds only for a particular choice of units. Such a choice is implicit in 
the choice of periodicity of ${\underline x}$. If the period is chosen to be $R_B$, which can be interpreted as the radius of the compact dimension,
one finds that the unit of quantization of $\tilde m$ is $1/R_B$. 
That is\foot{"Note added in proof: we have implicitly assumed that
the IIB string coupling constant $g_B$ is unity. As recently shown 
[\NEW] the right hand side should be replaced by $n/g_BR_B$ 
when $g_B\ne1$; the T-duality transformation between $g_B$ and the IIA 
string coupling constant $g_A$ then leads to a quantization condition 
of the form $m\sim n/(g_A\sqrt{\alpha'})$ in which the IIA mass parameter
is expressed entirely in IIA terms.},
$$
\tilde m = {n\over R_B}
\eqn\iifourb
$$
for integer $n$. Recall now that the equivalence of the 7-brane 
with the 8-brane under T-duality requires that $m=\tilde m$. This means, assuming IIB U-duality, that the IIA 8-brane solution can be mapped to 
a IIB 7-brane solution by T-duality only if the cosmological constant 
$m$ of the massive IIA theory is quantized as above, i.e. each time 
one passes through a IIA 8-brane the cosmological constant must jump 
by an integer multiple of basic unit $1/R_B$.

The single 8-brane solution should be related  
to the
Dirichlet 8-brane of [\pol]. This is a string background in which  
open string
states arise with fixed (Dirichlet) boundary conditions that are  
imposed in one
space-like dimension at one or both ends of the string. These  
conditions
restrict at least one of the end-points of open strings to lie in the
nine-dimensional worldvolume of an 8-brane. The 8-brane couples to a  
9-form
gauge field with a ten-form field strength $F_{10}$. If the new IIA
supergravity constructed here is indeed the effective field theory of  
the IIA
superstring in the presence of this 10-form field strength then it  
should be
possible to recover the Lagrangian \nseven\ by string theory  
considerations.
Neglecting terms of order $B^2$, which in any case follow from gauge
invariance, the only term in \nseven\ that is linear in $F_{10}$ is
proportional to
$$
(\varepsilon F_{10}) dA\cdot B\ .
\eqn\kone
$$
This is the crucial term that has to be reproduced in string theory.
There is a vertex operator in the RR sector of the type IIA theory  
that
couples a ten-form field strength to the worldsheet. This vertex  
operator has
the form $F_{10}{\bar S}S$, where $S$ is the spacetime spinor  
worldsheet field
of the spacetime supersymmetric worldsheet action. There are  
non-trivial tree
diagrams that mix $F_{10}$ with fields from the RR and NSNS sectors,  
producing
a term of the form \kone, as required. The requirements of gauge  
invariance
suggest that a more systematic consideration of string theory in the  
presence
of D-branes would produce the full effective Lagrangian \nseven.

Since all the $p$-brane solutions of D=10 IIA supergravity for $p<8$  
can be
viewed as arising from some 11-dimensional supermembrane theory,  
or `M-theory',
[\BST,\PKT,\EW,\JHS,\HW] it would be surprising if the 8-brane did  
not also
have an 11-dimensional interpretation. The obvious possibility is  
that the D=10
8-brane is the double-dimensional reduction of a D=11 supersymmetric  
9-brane.
Such an object would be expected (see [\PKTb]) to carry a 9-form
`charge' appearing in the D=11 supertranslation algebra as a central  
charge.
This is possible because the 2-form charge normally associated with  
the D=11
supermembrane is algebraically equivalent to a 9-form. It is not easy  
to see
how to implement this idea, however, since there is no `massive' D=11
supergravity theory. One possibility is suggested by the recent  
interpretation
[\HW] of the heterotic string as an $S^1/\Z_2$ compactified M-theory.  
Since the
compactification breaks half the supersymmetry and the compactifying  
space is
actually the closed interval, the two D=10 spacetime boundaries
might be viewed as the worldvolumes of two D=11 9-branes.

Less ambitiously, one could try to relate the massive IIA  
supergravity theory
to D=11 supergravity via some lower dimension, in the same way that  
the IIB
theory is related to it via reduction to D=9. In fact, this can be  
done by
compactification to D=8. To see this we first observe that there is  
clearly a
new massive N=2 D=8 supergravity theory obtainable {\it either} from  
the
massive N=2 D=9 theory (by the same procedure used to obtain the  
latter from
the massive IIA theory in D=10) {\it or} from the massless N=2 D=9  
theory by
Scherk-Schwarz dimensional reduction (using the global $U(1)$  
symmetry
inherited from the D=10 IIB theory). Thus, solutions of this massive  
N=2 D=8
supergravity theory should be liftable to D=9 as solutions of {\it  
either} the
massless {\it or} the massive N=2 D=9 supergravity theory. However,  
solutions
of the latter are also solutions of the massive D=10 IIA theory while  
solutions
of the former are also solutions of D=11 supergravity.

In light of this we may now ask to what solution of D=11 supergravity  
does the
8-brane solution of the massive IIA theory correspond? According to  
the above
procedure we should first double dimensionally reduce the 8-brane to  
D=8, where
it can be interpreted as a 6-brane. This 6-brane solution can of  
course be
lifted to D=9 as a 7-brane solution of the massive N=2 D=9  
supergravity theory,
but we expect that it can also be lifted to D=9 as a 6-brane solution  
of the
{\it massless} N=2 D=9 theory which can then be lifted to D=10 as the  
6-brane
solution of the massless IIA theory\foot{It can also be considered as  
a 7-brane
solution of the $S^1$-compactified IIB theory.}. As shown in [\PKT],  
this
6-brane solution is a non-singular solution of D=11 supergravity  
analogous to
the D=5 KK monopole. Thus the M-theory interpretation of the massive  
IIA
8-brane would appear to be as the KK 6-brane, at least on  
compactification from
D=11 to D=8.

\vskip 1cm
\centerline{\bf Acknowledgements}
\vskip 0.5cm
We would like to thank A.~Ach\'ucarro, J.H. Schwarz and K.S. Stelle for  
discussions.
G.P. would like to thank G. 't Hooft and B. de Wit for an invitation  
to visit
the university of Utrecht, and the particle physics group of the  
university of
Groningen for their hospitality.
E.B. would like to thank DAMTP for its hospitality.
G.P. is supported by a University Research Fellowship from the
Royal Society. The work of E.B. has been made
possible by a fellowship of the Royal Netherlands Academy of Arts and
Sciences (KNAW).

\vskip 1cm
\centerline{APPENDIX: {\bf An $SL(2,\R)$--formulation of $D=10$
Type IIB Supergravity}}
\vskip 0.5cm

In section 5 we used the supersymmetry transformations
of type IIB supergravity in ten dimensions in a rather simple
form. In this appendix we present the relation of our formulation
to that given in [\Schwarz]. Since we make no reference to D=9  
fields in this
section we shall drop the hats on the D=10 fields.

The scalars of IIB supergravity parametrize the
$SU(1,1)/U(1)$ coset. They form an $SU(1,1)$ doublet,
and under the local $U(1)$ they transform,
with weight $-1$, as follows:
$$
       \phi'_\alpha = \exp(-i\Lambda)\phi_\alpha\, .
\eqn\aaone
$$
Here $\alpha=1,2$ and $\phi_\alpha$ satisfies
 $|\phi_1|^2-|\phi_2|^2=1$.
The complex fermions $\psi_\mu$ and $\lambda$ have $U(1)$ weights  
$-1/2$ and
 $-3/2$ respectively.

The first step, also worked out in [\Schwarz],
is to fix a U(1) gauge by chosing $\phi_1$ real:
$$
\eqalign{
  \phi_1 = (\phi_1)^* &= {1\over \sqrt{1-\Phi^*\Phi}}\,,\cr
  \phi_2 &= {\Phi\over \sqrt{1-\Phi^*\Phi}}\, .
}
\eqn\aatwo
$$
This gauge choice is not invariant under $SU(1,1)$ or supersymmetry,
which requires
redefinition of these symmetries  with compensating $U(1)$  
transformations.
The complex field $\Phi$ can be written as
$$
     \Phi(x) \equiv {1+ i\tau(x)\over 1-i\tau(x)},\qquad
     \tau(x) \equiv \ell(x)+ie^{-\varphi(x)}\,.
\eqn\aathree
$$
The $SL(2,\R)$ transformations of $\tau$ are now given by:
$$
   \tau'  = {c + d\tau\over a + b\tau}\,,\qquad ad-bc=1\,.
\eqn\aafour
$$

In [\BHO,\BJO] an action for the bosonic part of IIB supergravity
was given in terms of the real scalars $\ell$ and $\varphi$, where
$\varphi$ was identified as the dilaton.
Even though this bosonic action is simple, supersymmetry still
is quite complicated if no further redefinitions are made.
In particular, the variation of the gravitino still contains
the composite U(1)-gauge field $Q_\mu$, which is a complicated function
of $\tau$. Also, the transformation rule of $\lambda$ is
nonlinear.

The following redefinitions of the Type IIB fermions
simplifies matters considerably:
$$
\eqalign{
   \tilde \psi_\mu &= \exp{(-\half i\theta(x))}\,\psi_\mu\,,\cr
   \tilde \epsilon &= \exp{(-\half i\theta(x))}\,\epsilon\,,\cr
    \tilde\lambda  &= \exp{(-\tfrac{3}{2}i\theta(x))}\,\lambda\, , }
\eqn\aafive
$$
where the function $\theta(x)$ is defined by
$$
    \exp{(-2i\theta(x))} = {1-i\tau\over 1+i\tau^*}\,.
\eqn\aasix
$$
The $SL(2,\R)$ transformations of $\tilde \psi, \tilde\epsilon,
\tilde\lambda$ are as follows:
$$
\eqalign{
\tilde \lambda^\prime &= \bigg ( {a+b\tau^*\over a+b\tau}
\bigg )^{\tfrac{3}{4}}\tilde\lambda\, ,\cr
\tilde \epsilon^\prime &= \bigg ( {a+b\tau^*\over a+b\tau}
\bigg )^{\tfrac{1}{4}}\tilde\epsilon\, ,\cr
\tilde \psi^\prime &= \bigg ( {a+b\tau^*\over a+b\tau}
\bigg )^{\tfrac{1}{4}}\tilde\psi\, .}
\eqn\aasixa
$$
Note that the redefined fermions are invariant under the abelian  
subgroup of
$Sl(2;\R)$ defined by setting $a=d=1$, $b=0$. This subgroup, which  
acts on
$\tau$ as $\tau'= \tau +c$, is the group used for SS dimensional  
reduction in
section 5.

The redefinition \aafive\ leads to the following simplified, Einstein  
frame,
supersymmetry transformations (now omitting the tildes):
$$
\eqalign{
   \delta e_\mu^a &= \half \,\bar\epsilon\,
    \Gamma^a \psi_\mu + {\rm h.c.}\,,\cr
   \delta\psi_\mu &= {\cal D}_\mu\epsilon
       + \tfrac{i}{4}\epsilon \,e^\varphi\partial_\mu \ell
       -\tfrac{i}{192}\Gamma^{(5)}\Gamma_\mu\epsilon \,F_{(5)}\cr
 &  -\tfrac{1}{16}\left(\Gamma_\mu\Gamma^{(3)}+2\Gamma^{(3)}
\Gamma_\mu\right)
   \epsilon^* e^{\varphi/2}\left( H^{(1)}-\ell H^{(2)}
         -ie^{-\varphi}H^{(2)}\right)_{(3)}\,,\cr
   \delta A_{\mu\nu\lambda\rho} &=
  i\,\bar\epsilon\, \Gamma_{[\mu\nu\lambda}\psi_{\rho]} + {\rm h.c.}
 -6 \epsilon_{ij}B^{(i)}_{[\mu\nu}\delta B^{(j)}_{\lambda\rho]}\,,\cr
 \delta B^{(1)}_{\mu\nu} &=
  \half\left(e^{-\varphi/2} + i\ell e^{\varphi/2}\right)
  \left(\bar\epsilon^*\Gamma_{[\mu}\psi_{\nu]}
        -\half \,\bar\epsilon\, \Gamma_{\mu\nu}\lambda\right)
     + {\rm h.c.}\,,\cr
 \delta B^{(2)}_{\mu\nu} &=
    \tfrac{i}{2} \,e^{\varphi/2}
   \left(\bar\epsilon^*\Gamma_{[\mu}\psi_{\nu]}
        -\half \,\bar\epsilon\, \Gamma_{\mu\nu}\lambda\right)
   + {\rm h.c.} \,,\cr
 \delta\lambda &=
     \tfrac{1}{4}\Gamma^\mu\epsilon^*
   \left( \partial_\mu\varphi +ie^\varphi\partial_\mu \ell
\right)\cr
  &\qquad +\tfrac{1}{8}\Gamma^{(3)}\epsilon \,e^{\varphi/2}
   \left(H^{(1)} - \ell H^{(2)} -ie^{-\varphi}
 H^{(2)}\right)_{(3)}\,,\cr
  \delta \ell &= i{\rm
 e}^{-\varphi}\,\bar\epsilon\,\lambda^* + {\rm h.c.}\,,\cr
  \delta\varphi &= \,\bar\epsilon\,\lambda^* + {\rm h.c.}     \,.
}
\eqn\aaseven
$$
The covariant derivative ${\cal D}\epsilon$ in the variation of the  
gravitino
contains only the gauge field of local Lorentz transformations but  
no
composite $U(1)$ gauge field $Q_\mu$. Due to the redefinitions
of the fermions
the only remnant of $Q_\mu$ is a single
$e^\varphi\partial \ell$ term. The three-forms $H^{(i)},\ i=1,2$ are  
the field
strengths of
the two-form gauge fields $B^{(i)}$.
The field strength $F_{(5)}$ satisfies a self-duality
condition, and is given by
$$
   F_{\mu\nu\lambda\rho\sigma} \equiv
  \partial\,{}_{[\mu} A_{\nu\lambda\rho\sigma]}
     -6\epsilon_{ij}B^{(i)}_{[\mu\nu}H^{(j)}_{\lambda\rho\sigma]}\,.
\eqn\aaeight
$$
The above transformation rules should still be completed with
terms bilinear in the fermion fields. In
principle these can be constructed from the results of [\Schwarz],  
using the redefinitions given above.

\refout

\end